# He-CF4-CH4 ternary mixtures as target gas for the CYGNO directional dark matter experiment

F.D. Amaro [1], R. Antonietti [2,3], E. Baracchini [4,5], L. Benussi [6], C. Capoccia [6], M. Caponero [6,7], L.G.M de Carvalho [8], G. Cavoto [9,10], I.A. Costa [6], A. Croce [6], M. D'Astolfo [4,5], G.D'Imperio [10], G. Dho [6], E. Di Marco [10], J.M.F. dos Santos [1], D. Fiorina [4,5], F. Iacoangeli [10], Z. Islam [4,5], H. P. Lima Jr [4,4], G. Maccarrone [6], R.D.P. Mano [1], D. J. G. Marques [4,5], L.G.M. de Carvalho [8], G. Mazzitelli [6], P. Meloni [2,3], A. Messina [9,10], C.M.B. Monteiro [1,*], R.A. Nobrega [8], I.F. Pains [8], E. Paoletti [6], F. Petrucci [2,3], S. Piacentini [4,5], D. Pierluigi [6], D. Pinci [10], F.Renga [10], R.J. da C. Roque [1], A. Russo [6], G.Saviano [6,11], P.A.O.C. Silva [1], R.C. Sousa [1], N.J.C. Spooner [12], R. Tesauro [6], S. Tomassini [6] and D. Tozzi [9,10]

[1]LIBPhys; Department of Physics; University of Coimbra; 3004-516 Coimbra; Portugal
[2]Dipartimento di Matematica e Fisica; Universit`a Roma TRE; 00146; Roma; Italy
[3]Istituto Nazionale di Fisica Nucleare; Sezione di Roma Tre; 00146; Rome; Italy
[4]Gran Sasso Science Institute; 67100; L'Aquila; Italy
[5]Istituto Nazionale di Fisica Nucleare; Laboratori Nazionali del Gran Sasso; 67100; Assergi; Italy
[6]Istituto Nazionale di Fisica Nucleare; Laboratori Nazionali di Frascati; 00044; Frascati; Italy
[7]ENEA Centro Ricerche Frascati; 00044; Frascati; Italy
[8]Universidade Federal de Juiz de Fora; Faculdade de Engenharia; 36036-900; Juiz de Fora; MG; Brasil
[9]Dipartimento di Fisica; Universit`a di Roma Sapienza; 00185; Roma; Italy
[10]Istituto Nazionale di Fisica Nucleare; Sezione di Roma; 00185; Roma; Italy
[11]Dipartimento di Ingegneria Chimica; Materiali e Ambiente; Sapienza Universit`a di Roma; 00185; Roma; Italy
[12]Department of Physics and Astronomy; University of Sheffield; Sheffield; S3 7RH; UK

# Abstract

The CYGNO collaboration is advancing a high-resolution optical Time Projection Chamber (TPC) for directional dark matter searches and solar neutrino spectroscopy at LNGS. The detector uses a He-40%CF4 gas mixture at atmospheric pressure and a triple-GEM cascade for ionization signal amplification. Scintillation light from GEM electron avalanches is read out using sCMOS cameras, enabling high sensitivity to interactions in the few keV range, alongside precise ionization event tracking and particle identification. This study investigates the effects of adding 3–10% methane to the He-40%CF4 mixture. Methane improves the electrical stability of the TPC, allowing for higher GEM voltages before discharge onset, which compensates for its scintillation quenching and leads to enhanced overall scintillation yield. Compared to prior studies performed with isobutane, methane demonstrates a smaller quenching effect on visible and UV photons while maintaining good energy resolution. Importantly, the inclusion of methane lowers the dark matter detection threshold by providing a lighter target and extending track lengths of light nuclear recoils, thus enhancing directional discrimination. These results establish methane as a promising additive for optimizing CYGNO's performance in detecting low-mass dark matter candidates.

* Corresponding author cristinam@uc.pt

# Introduction

Weakly Interacting Massive Particles (WIMPs) remain among the most compelling candidates for cold dark matter (DM), and numerous experiments are dedicated to their detection. WIMPs interact exclusively through gravitational and faint forces, making their detection challenging. Direct searches aim to measure their interactions with atomic nuclei by detecting ionization signals produced by nuclear or electronic recoils. Due to the extremely low interaction rate of WIMPs, detection requires large sensitive volumes with low thresholds. Additionally, it demands stringent background suppression, achieved by operating detectors underground and using materials with carefully minimized radioactivity [1-3]. Consequently, a severe suppression of background events is necessary, involving underground operation and a drastic suppression of the natural radioactivity of the detector materials through rigorous selection [4,5]. Directional sensitivity offers an additional method for background suppression by leveraging the motion of WIMPs relative to Earth. This enables discrimination between true WIMP interactions and isotropic background events, providing a unique signature for detection [6-12].

The CYGNO experiment [11-13] is developing a high-resolution optical Time Projection Chamber (TPC) for directional dark matter searches and solar neutrino spectroscopy. The ionisation signal is amplified by drifting the primary electrons towards a triple-Gas Electron Multiplier (GEM) stack, where they undergo charge avalanche. The secondary scintillation produced during charge avalanches is key to event detection, as it provides a clear optical signal that is recorded using a high-granularity sCMOS camera. Photomultiplier tubes (PMT) complement the optical readout by capturing the scintillation time profile, offering additional temporal information about ionization events. Combining the information of both photosensors, the ionisation track extension along the drift direction and its inclination with respect to the amplification plane can be inferred, enabling the reconstruction of the 3D direction of the tracks [14, 15].

The TPC filling gas is a He/$CF_4$ mixture at atmospheric pressure, which enables high sensitivity to interactions in the few keV energy range together with good particle identification. A He-40% $CF_4$ mixture at atmospheric pressure was found to be the best choice for extending CYGNO sensitivity to GeV and sub-GeV WIMP masses for spin independent coupling and good tracking capability [11]: a target with low mass nuclei allows for longer track lengths, resulting in improved determination of their direction, hence increasing the directional sensitivity; $CF_4$ is chosen for its high scintillation yield in the visible region, matching the sCMOS camera sensitivity, and its scintillation has been studied in detail [11,12,16]. In addition, fluorine provides sensitivity to spin dependent WIMP-proton interactions.

The momentum transfer between a WIMP and the gas target is maximised when their masses are equal. Therefore, the addition of hydrogen, the lightest atom, to the CYGNO-TPC would further improve the detector sensitivity to low WIMP masses and lead to a gas target with lower density. It improves the tracking capabilities in terms of drift velocity, diffusion and longer track lengths [11], thus improving the directionality, particle identification and background rejection [13]. Although pure hydrogen cannot be easily added to the CYGNO-TPC due to experimental constraints, hydrocarbons such as isobutane, i-$C_4H_{10}$, and methane, $CH_4$, are good alternatives. Adding 2% of isobutane to He-40%$CF_4$ extends the WIMP mass sensitivity to 0.32 GeV/$c^2$ for spin-dependent and spin-independent coupling, whereas the base mixture is only sensitive down to 0.9 GeV/$c^2$ for spin-independent and 2 GeV/$c^2$ for spin-dependent coupling [11]. In a former publication [17] we have investigated the impact of the addition of 1 to 5% of isobutane to the He-40%$CF_4$ mixture in terms of the total number of electrons and scintillation photons produced in GEM avalanches as a function of GEM voltage. While the total number of avalanche electrons increases with increasing isobutane content due to Penning transfer from excited He atoms to isobutane molecules, contributing to increase the scintillation output, the total number of photons emitted in the avalanches decreases, nonetheless, due to quenching, by the isobutane molecules, of the excited state species resulting from He and $CF_4$ excitations [17]. The sensitivity improvement due to the presence of H atoms is expected to be similar for other H-rich molecules, e.g. $CH_4$. As shown in Figure 1, a 5% $CH_4$ admixture to the He-40%$CF_4$ mixture would extend the sensitivity of the CYGNO-TPC down to 0.4 GeV/c$^2$. Therefore, the optimum hydrocarbon additive to He-$CF_4$, one that improves detector sensitivity with minimum negative impact on the optical readout, should be found.

* Corresponding author cristinam@uc.pt

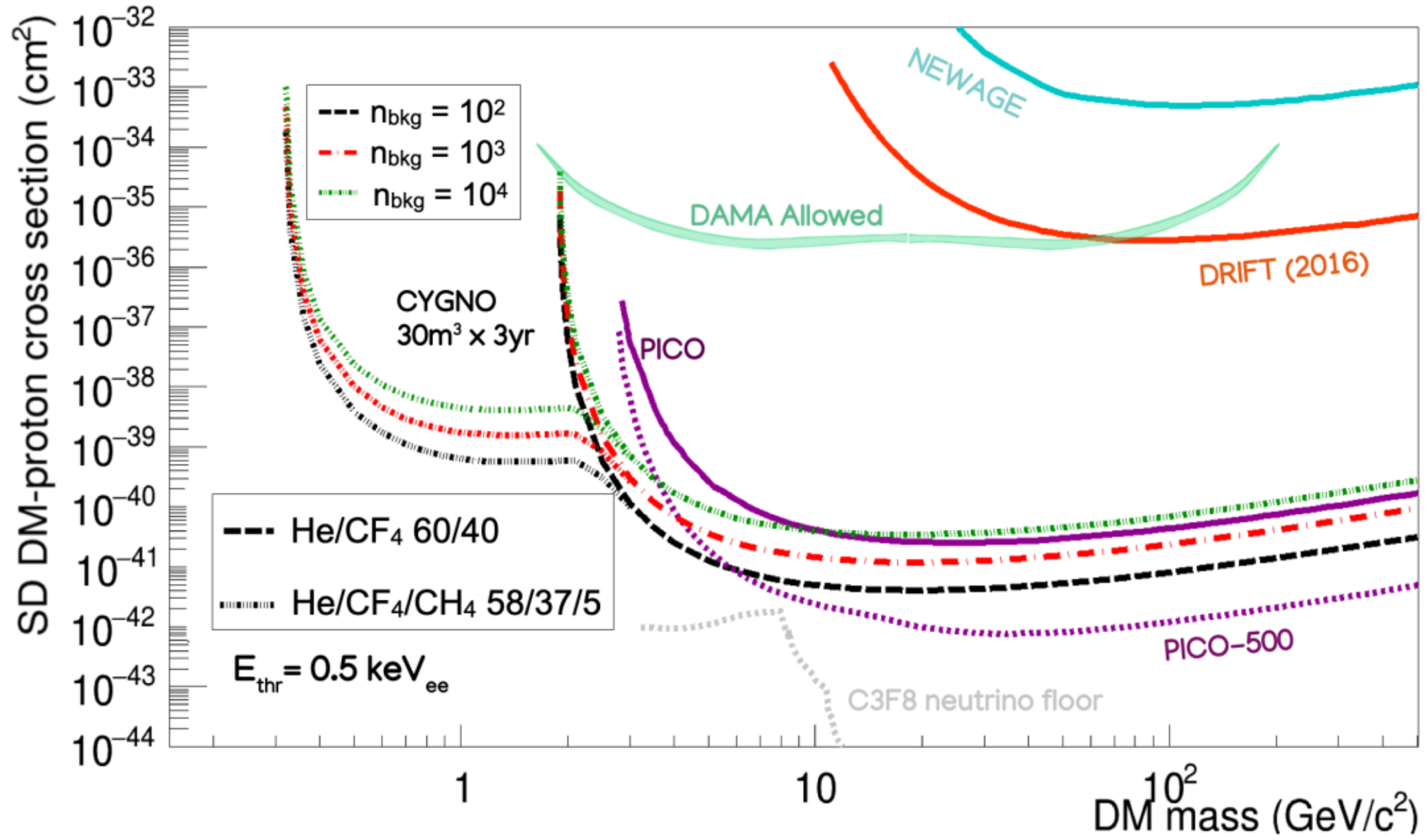

Figure 1: Spin-dependent sensitivity for WIMP-proton cross sections for the 30 m$^3$ CYGNO TPC with an operative threshold of 0.5 keV and for 3 years of exposure with different background level assumptions corresponding to different possible number of background events per year, $N_{bkg}$ = 100 (black), 1000 (red) and 10,000 (dark green). The dashed curves show the sensitivity for a He/CF4 (60/40) mixture, while the dotted curves show the sensitivity for a He/CF4/methane (58/40/5) mixture.

In this work we study the impact of $CH_4$ addition to the He-40%$CF_4$ mixture. The charge amplification and the scintillation yield will be studied as a function of GEM bias, i.e., the voltage difference between the bottom and top electrode, for different concentrations of methane, in the 3 to 10% range.

# Experimental setup

A dedicated setup was built to determine the absolute scintillation output from GEM avalanches. This experimental setup, previously employed in He-$CF_4$-isobutane studies [17], includes a GEM-based amplification system coupled to a Large Area Avalanche Photodiode (LAAPD) for scintillation readout, designed for high-precision absolute photon yield measurements. The schematic of the experimental setup is presented in Fig.2. A 6-mm deep drift region is delimited by the GEM top electrode and a stainless-steel mesh, 80-μm diameter wires, 900 μm in pitch, while the GEM bottom electrode and a second stainless-steel mesh delimit the 2-mm thick induction region. Below the induction mesh an LAAPD, an Advanced Photonics Instruments Deep-UV series with a 16-mm diameter sensitive area [18], was used to read out the scintillation emitted from the GEM avalanches. The GEM was made at CERN with standard dimensions: a 50-μm thick Kapton foil with a 5-μm thick Cu coating on both sides and perforated with holes, 70-μm and 50-μm diameter at Cu and Kapton, respectively, in a hexagonal pattern. The meshes and GEM foil were glued to MACOR frames using low-outgassing, electrically insulating glue to prevent gas contamination and electrical discharges. Four Teflon pillars were used to hold the meshes and the GEM foil inside an aluminium chamber, sealed with Viton O-rings and equipped with Swagelok gas connectors and SHV electrical feedthroughs.

* Corresponding author cristinam@uc.pt

Throughout the present studies, the drift field was set to 500 V/cm and the induction field to 300 V/cm, oriented to promote the collection and readout of avalanche electrons at the bottom electrode of the GEM. CAEN N471 high voltage supplies, with a current limit set to 100 nA, were used to bias the meshes and the GEM electrodes through custom-made low-pass filters with an RC constant of 20 ms.

The chamber was operated in flow mode with a gas flow of about 4 $dm^3\ h^{-1}$, maintaining consistent gas composition and minimizing contamination during measurements. The flow controller is unable to maintain flows of methane below 0.1 $dm^3\ h^{-1}$, hence methane admixtures of 1% and 2% could not be evaluated with the present setup. The uncertainty in the concentrations is less than 0.25 %, 0.025 % and 0.25 %, in absolute value, for CH4, CF4 and He, respectively.

The emission spectrum of the He-40%$CF_4$ mixture is well known [16], Fig.3, being $CF_4$ the responsible for this emission. The LAAPD has a spectral sensitivity in the 120–1000 nm range [18], Fig.3. To study only the photon emission in the visible range, a borosilicate glass filter, N-B270 [19], was placed on top of the LAAPD.

* Corresponding author cristinam@uc.pt

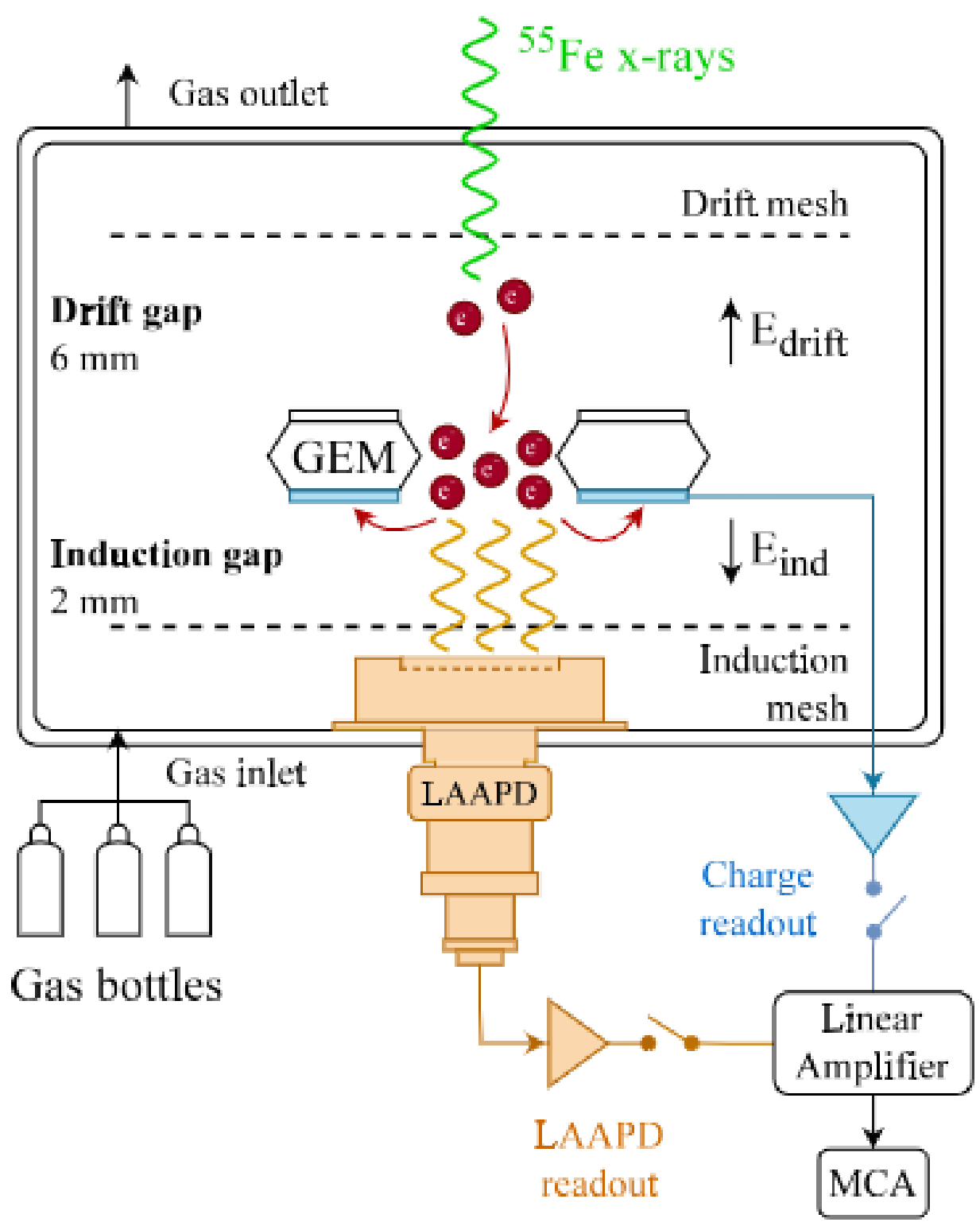

Figure 2: Schematic of the experimental setup used for the present studies.

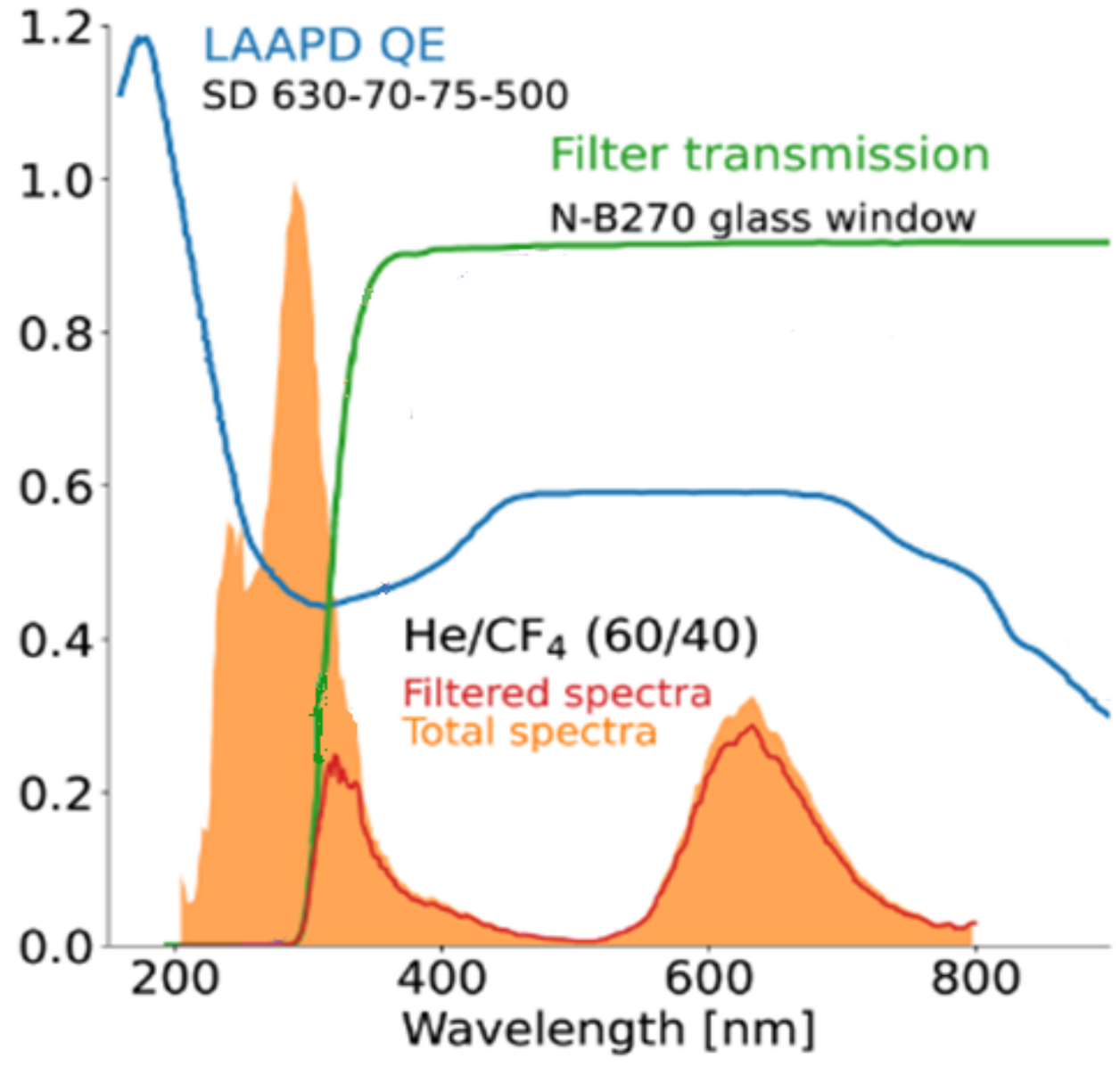

Figure 3: He-40%$CF_4$ emission spectrum (orange area), taken from [16]; LAAPD quantum efficiency (blue), taken from [18] and N-B270 borosilicate glass filter transmission (green) taken from [19]. The filtered spectrum (red) was computed by multiplying the N-B270 filter transmission with the total emission spectrum of He-40%$CF_4$.

* Corresponding author cristinam@uc.pt

# Methodology

A 2-mm collimated 5.9-keV x-ray beam was used to promote gas ionization in the drift region. The ionization electrons drifted toward the GEM holes, where the high electric field initiate electron avalanches, amplifying the primary signal for subsequent detection. Those avalanche electrons are collected at the GEM bottom electrode and a pulse-height distribution is recorded with a multichannel analyser through a pre-amplifier and a linear amplifier electronic chain. Simultaneously, the photons emitted from the electron avalanches are read out by the LAAPD and the pulse-height distributions at the LAAPD output are recorded with an analogous electronic chain, Fig. 2.

The combined use of x-rays with an LAAPD photosensor allows for a straightforward method to determine the absolute scintillation yield produced in the gas [20-23], eliminating the need for complex calibration procedures. In the LAAPD, alongside the detection of the scintillation pulses produced in the GEM charge avalanches, resulting from the 5.9-keV x-ray interactions in the target gas, a part of the 5.9-keV x-rays is transmitted through the gas volume and interacts directly in the LAAPD sensitive region. Consequently, the pulse-height distributions recorded at the output of the LAAPD features two peaks, one due to the scintillation pulses and another one due to the 5.9-keV x-ray direct interactions in the LAAPD, as shown in Fig.4.

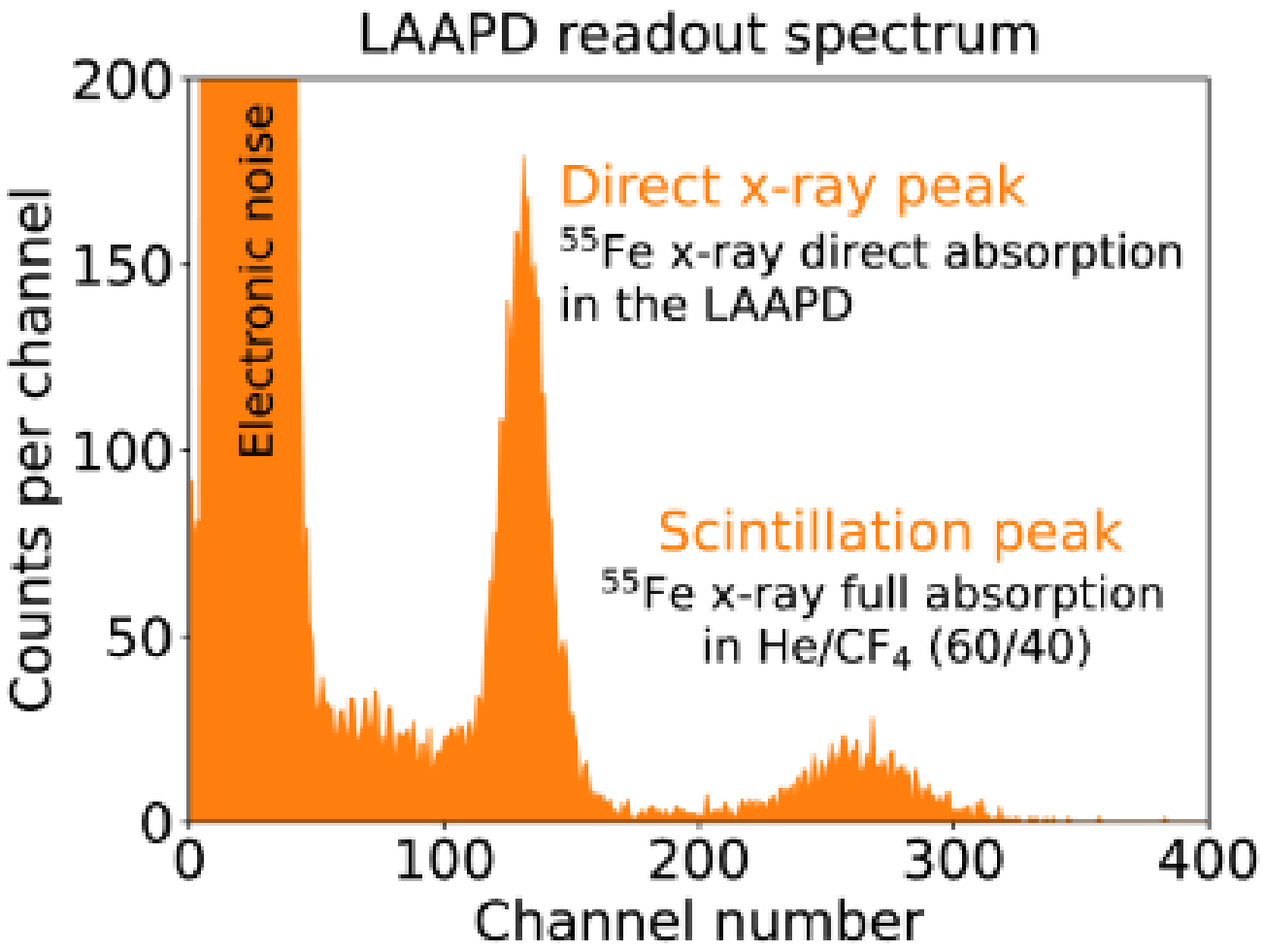

Figure 4: Typical pulse-height distribution obtained for 5.9-keV x-rays, and for a He-40%CF4 gas mixture, a drift field of 500 V/cm, an induction field of 300 V/cm and a GEM biasing of 520 V.

The Si *w*-value, the average energy required to produce an electron-hole pair, is known to be 3.62 eV [24]; the centroid of the direct x-ray interaction peak, $C_x$, can be used to calibrate the Multi-Channel Analyser (MCA) channel in terms of electron-hole pairs. In this case,

$$Cal = 5.9\ keV/C_x/3.62\ eV \quad \text{(electron-hole pair/channel)}.$$

* Corresponding author cristinam@uc.pt

From the ratio of the scintillation-peak-to-direct-x-ray-interaction-peak, $C_{sc}/C_x$, the average number of electron-hole pairs produced in the LAAPD by the gas scintillation emitted in the electron avalanches resulting from 5.9 keV interactions in the drift region can be determined as

$$N_{APD} = (C_{sc}/C_x) \times (5.9\ \text{keV}/3.62\ \text{eV}) \qquad \text{(electron-hole pair)}.$$

Being the average quantum efficiency of the LAAPD, $Qe_{av}$, known then the number of photons hitting the LAAPD can be calculated and, correcting for the fraction of the solid angle subtended by the LAAPD relative to the scintillation region, $\Omega_{rel}$, the average total number of photons emitted in the avalanches resulting from the interactions of 5.9-keV x-rays in the drift region will be

$$N_{ph} = N_{APD}/(Qe_{av} \times \Omega_{rel} \times T) = (C_{sc}/C_x) \times (5.9\ \text{keV}/3.62\ \text{eV})\ /(Qe_{av} \times \Omega_{rel} \times T) \qquad \text{(photons)},$$

where T is the mesh optical transmission.
Additionally, we can define the scintillation yield produced in the gas avalanches as the number of photons emitted per unit of energy deposited in the gas drift volume,

$$Y_{keV} = N_{ph}/5.9\ \text{keV} = (C_{sc}/C_x) \times (1000/3.62)\ /(Qe_{av} \times \Omega_{rel} \times T) \qquad \text{(photons/keV)}.$$

The number of avalanche electrons collected at the GEM bottom electrode is determined from the centroid channel of the respective pulse-height distribution, after the MCA calibration in terms of number of electrons per channel. This was achieved using a calibrated capacitor and a precision pulse generator.

# Experimental results

### Charge readout

Figure 5 depicts the average number of avalanche electrons as a function of GEM biasing voltage, for different methane percentage additions to the base-mixture He-40%$CF_4$. For each gas mixture the GEM bias was increased until the onset of discharges, i.e., more than one current spike above 100 nA per minute. The results demonstrate the effect of methane addition on charge amplification.

* Corresponding author cristinam@uc.pt

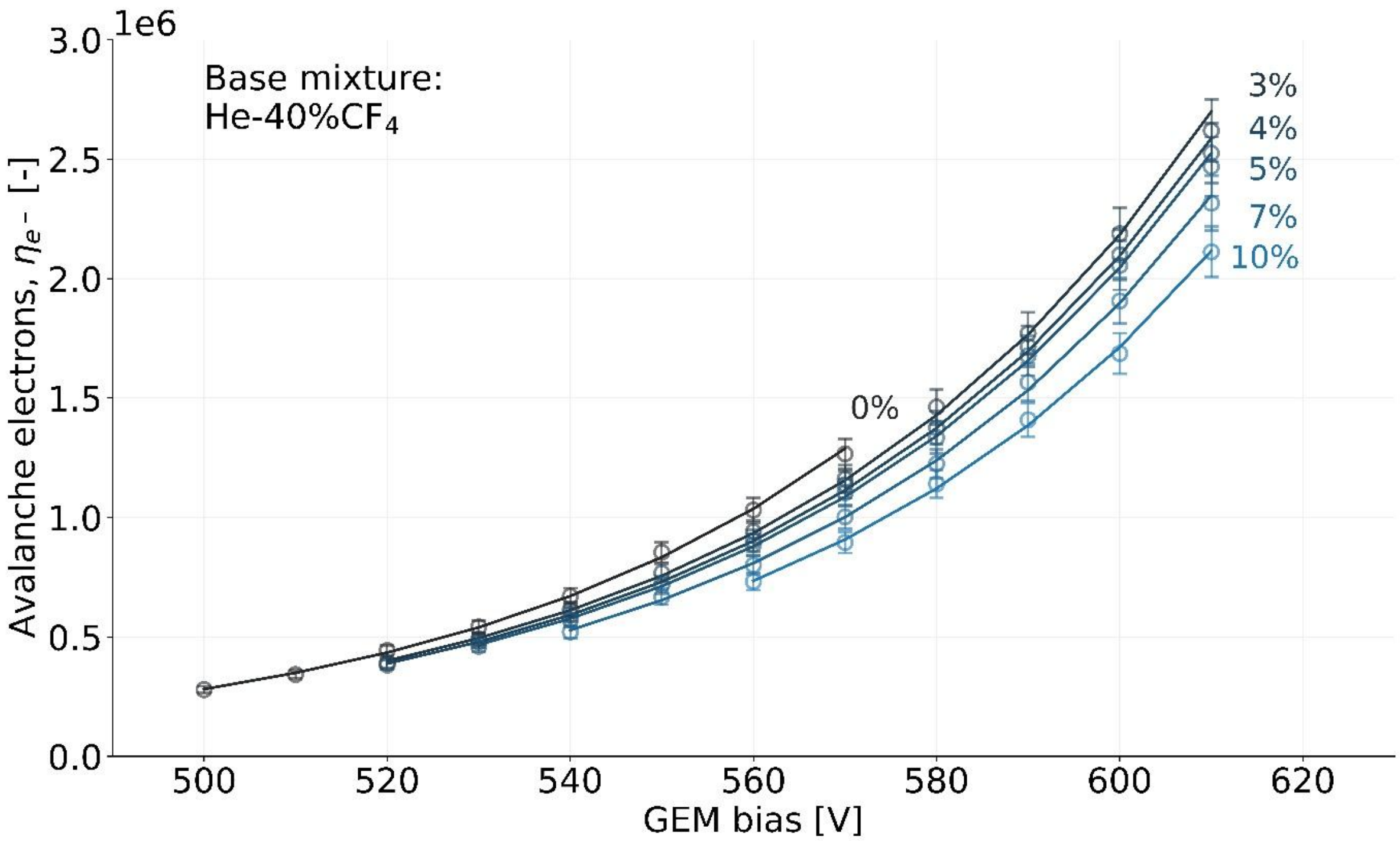

Figure 5: GEM-induced charge amplification for the different methane percentage additions to He-40%$CF_4$: the measurements are depicted with open circles, whereas the solid lines show the corresponding exponential trendlines. The percentages correspond to the amount of methane added to He-40%$CF_4$.

In opposition to what we observed for isobutane admixtures, increasing the methane content decreases the number of avalanche electrons for the same GEM biasing voltage. Although the ionization potential of methane, 12.99 eV [25], is still below the energy of the metastable states of He, rendering it eligible for Penning transfer mechanism, the obtained results show that there is no additional charge produced. One premise is that He excited states may still transfer their energy to vibrational and rotational states of $CH_4$ molecules without ionising them. Another hypothesis can be found in [26], where it is shown that the production of metastable states of He is much slower than the direct electronic ionization and excitation of $CH_4$, which may prevent an effective Penning energy transfer.

Also in opposition to isobutane, for the methane admixtures the GEM biasing voltage could be increased by 40 V relative to that of the base mixture, before the onset of discharges, suggesting that small amounts of methane increase the electrical stability of the He-40%$CF_4$/$CH_4$ mixture. Hence, He-40%$CF_4$/$CH_4$ could be a good ternary gas mixture for a long-running detector, such as the CYGNO-TPC.

Comparable to isobutane, the addition of small percentages of methane to He-40%$CF_4$ does not seem to degrade the energy resolution of the charge signals. For higher $VGEM$ values, the achieved energy resolution for $^{55}$Fe x-rays does not exhibit a significant dependence on the methane content, reaching values around 14%, similar to those achieved in isobutane admixtures and comparable to what is achieved with the base mixture He-40%$CF_4$.

* Corresponding author cristinam@uc.pt

## Optical readout

Figure 6 shows the average yield for the total scintillation and for the visible component regarding the different methane percentages added to He-40%$CF_4$ as a function of GEM biasing voltage. As expected, both the total and the visible scintillation yield increase exponentially with increasing voltage applied across the GEM, consistent with the behaviour of charge avalanche processes. For the same GEM biasing voltage, increasing methane content from 0 to 5 %, 7 % and 10 % reduces total and visible scintillation yields to 50 %, 40 % and 30 %, respectively, meaning that methane quenches the photons emitted by the He-40%$CF_4$ mixture. The visible component constitutes approximately 60-66% of total scintillation for all methane admixtures, indicating that methane quenches visible and UV photons equally, within uncertainties.

The maximum scintillation yield that can be achieved with methane admixtures up to 7% content is higher than the one obtained for He-40%$CF_4$. This is a consequence of the increased electrical stability achieved with the introduction of methane, allowing higher voltages to be applied to the GEM before the onset of discharges. At high GEM biasing the minimum energy resolution obtained for all the mixtures is around 20%, for the total scintillation as well as for the visible component.

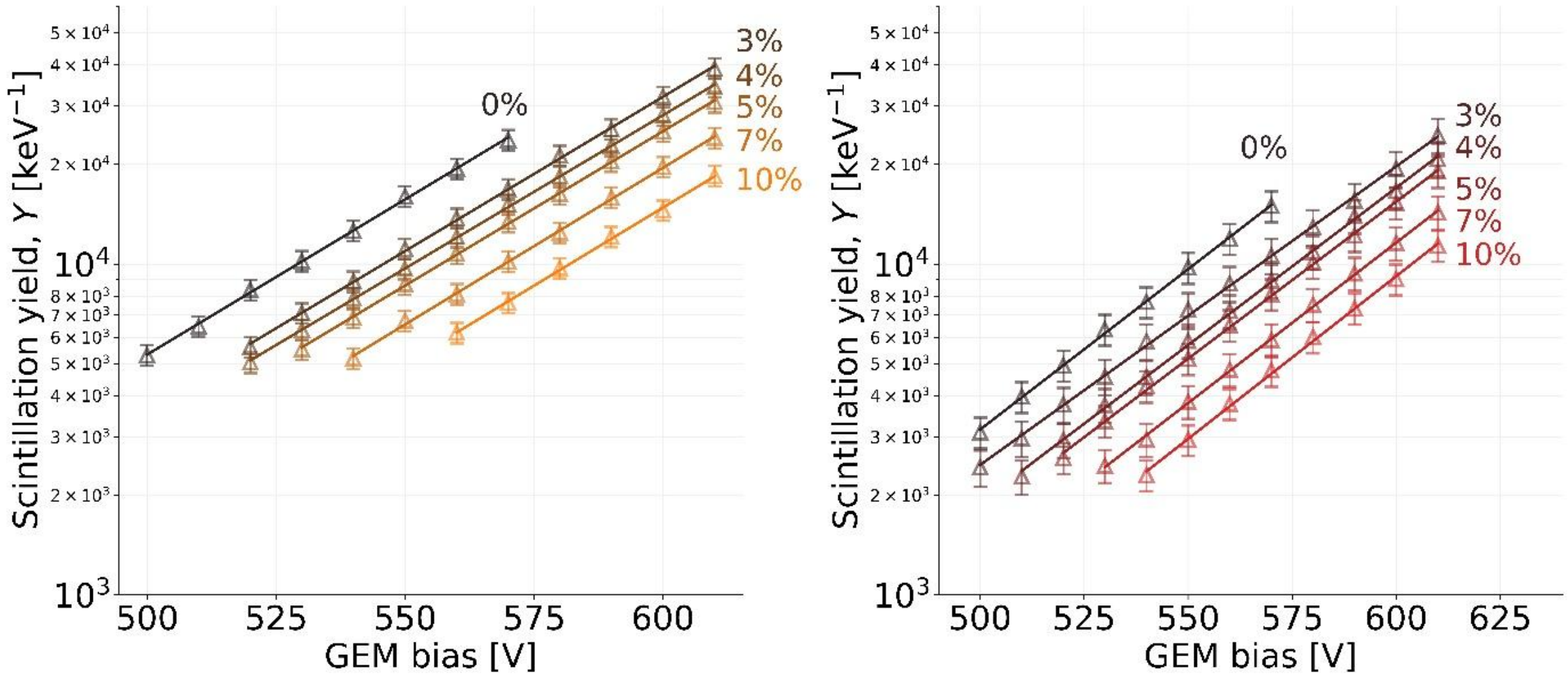

Figure 6: GEM-induced scintillation yield for different percentages of methane added to the base-mixture of He-40%$CF_4$: the total scintillation yield is presented on the left panel, whereas the yield for the visible component is presented on the right panel. The measurements are depicted with open markers and the solid lines illustrate the corresponding exponential trendlines. The percentages specify the amount of methane added to He-40%$CF_4$.

Figure 7 shows the average number of photons per avalanche electron as a function of GEM biasing voltage for admixtures with different methane content and for the total scintillation. As obtained for other gas mixtures [17], the number of photons emitted per avalanche electron is independent from GEM biasing and decreases with the amount of methane content in the mixture.

* Corresponding author cristinam@uc.pt

Table 1 summarizes the main results obtained for signal amplification in charge and in scintillation and for the several studied admixtures of He-40%$CF_4$/$CH_4$. The growth factor of the exponential fit to the charge and/or scintillation increase with increasing GEM biasing. Notably, methane admixtures achieve improved electrical stability while maintaining comparable energy resolution, but at the cost of reduced photon yield per avalanche electron.

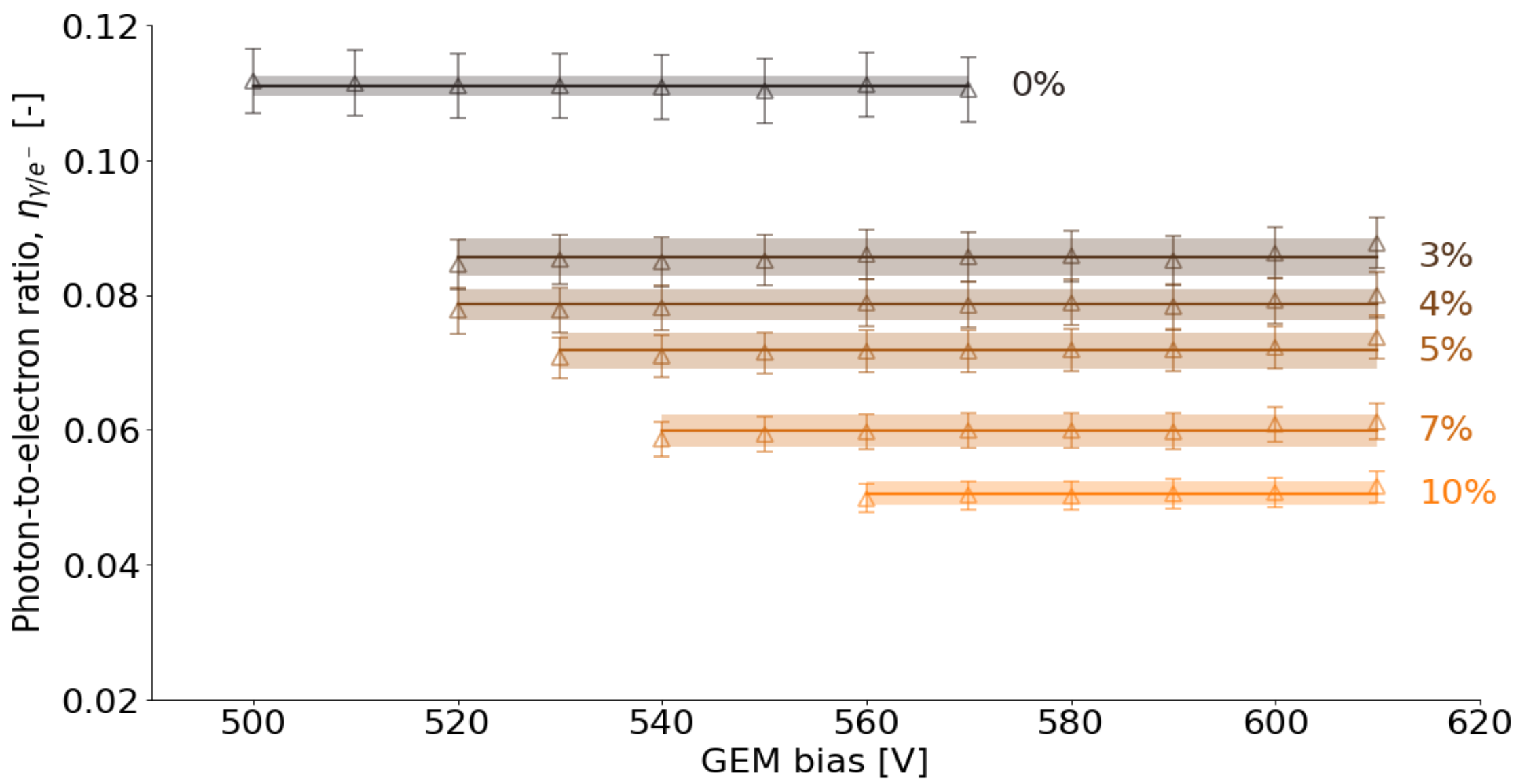

Figure 7: GEM-induced photon-to-electron ratio for the different admixtures of He-40%$CF_4$/$CH_4$: the total scintillation yield measurements are indicated in open triangles, whereas the horizontal lines and corresponding colour-areas represent the average value obtained for each methane content with the corresponding 3-σ confidence level. The percentages express the amount of methane added to the He-40%$CF_4$ base-mixture.

Table 1: GEM-induced signal amplification for the different percentages of methane additions to He-40%$CF_4$: the growth factors, β, maximum amplification, $\eta_{e^-}^{max}$ for avalanche electrons and $Y^{max}$ for the scintillation yield as well as minimum energy resolution, $FWHM^{min}$, are presented for both charge and optical readout. The photon-to-electron ratio for the total scintillation, $\gamma/e^-$, and the ratio of the visible component of the scintillation are also presented.

* Corresponding author cristinam@uc.pt

| | **0%** | **3%** | **4%** | **5%** | **7%** | **10%** |
|---|---|---|---|---|---|---|
| **Avalanche electrons** | | | | | | |
| $\beta$ | 0.0217(8) | 0.0212(5) | 0.0211(6) | 0.0211(6) | 0.0213(8) | 0.0212(12) |
| $\eta_{e^-}^{\max}$ ($\times 10^6$) | 1.27(6) | 2.62(13) | 2.53(13) | 2.47(12) | 2.32(12) | 2.11(11) |
| FWHM$^{\min}$ [%] | 13.48(19) | 13.85(19) | 14.11(20) | 14.36(21) | 14.20(17) | 14.17(19) |
| **Total scintillation** | | | | | | |
| $\beta$ | 0.0216(11) | 0.0215(8) | 0.0213(8) | 0.0215(9) | 0.0218(11) | 0.0217(17) |
| $Y^{\max}$ ($\times 10^4$) | 2.37(17) | 3.90(28) | 3.43(24) | 3.09(22) | 2.41(17) | 1.85(13) |
| $\gamma/e^-$ | 0.1110(34) | 0.0856(23) | 0.0785(23) | 0.0717(21) | 0.0599(18) | 0.0505(18) |
| FWHM$^{\min}$ [%] | 20.1(4) | 18.89(31) | 19.55(24) | 20.26(35) | 20.3(4) | 20.99(35) |
| **Visible scintillation** | | | | | | |
| $\beta$ | 0.0223(16) | 0.0208(10) | 0.0219(11) | 0.0219(12) | 0.0223(15) | 0.0226(16) |
| $Y^{\max}$ ($\times 10^4$) | 1.50(16) | 2.45(30) | 2.08(24) | 1.91(21) | 1.43(16) | 1.14(12) |
| FWHM$^{\min}$ [%] | 20.09(31) | 19.58(31) | 18.93(13) | 20.28(28) | 22.20(27) | 21.8(4) |
| Ratio [%] | 60.3(17) | 62.5(29) | 63(4) | 63(5) | 60(5) | 66(6) |

# Discussion

This section compares the effects of methane and isobutane [17] additions to the He-40%$CF_4$ base mixture, with a focus on maximum scintillation yields and photon-to-electron ratios. These comparisons aim to identify the optimal hydrocarbon additive for CYGNO's performance requirements. We define the hydrogen-equivalent concentrations of methane and isobutane admixtures in order to compare admixtures that attain the same WIMP detection efficiency, i.e. the concentration for which methane and isobutane have the same number of available H-atoms to interact with incoming WIMPs. We will also discuss the quenching mechanism of methane and isobutane regarding the scintillation of $CF_4$-based mixtures.

Figure 8 shows the maximum total and visible scintillation yields, obtained before the onset of discharges for methane and isobutane admixtures to He-40%$CF_4$. As shown, methane admixtures achieve higher maximum scintillation yields than isobutane at hydrogen-equivalent concentrations, emphasizing its advantages in optimizing optical performance. The results show that isobutane admixtures decrease the maximum scintillation yield that can be obtained with the base mixture. For 1 % isobutane, the maximum scintillation yield is 80 % of that achieved for the base-mixture He-40%$CF_4$, whereas for 5 % isobutane it decreases to 40 %. This substantial decrease could compromise signal-to-noise ratio in the CYGNO-TPC, limiting optical readout efficiency in long-term operations.

For methane concentrations below 7 % the scintillation yield of the He-40%$CF_4$ base mixture can be surpassed, reaching 1.6-fold above the one achieved for He-40%$CF_4$ with 3 % of methane. The increase in the maximum scintillation yield relative to the base mixture is achieved due to the electrical stability provided by methane, which increases the GEM biasing threshold for discharges. For 1 % and 2 % methane, the maximum scintillation yield should be even higher, but it was impossible to evaluate these concentrations due to the intrinsic resolution of the used flow controller.

The maximum scintillation yield of all methane admixtures is higher than the what was achieved for the hydrogen-equivalent concentrations of isobutane. This means that the optical gain of the

* Corresponding author cristinam@uc.pt

CYGNO-TPC can be increased with methane admixtures without losing efficiency to WIMP interactions.
Figure 9 shows the average number of photons emitted per avalanche electron as a function of methane and isobutane concentrations added to He-40%$CF_4$. Although both hydrocarbons quench the scintillation produced by He-40%$CF_4$, methane admixtures quench less scintillation photons per avalanche electron than the corresponding hydrogen-equivalent isobutane admixture.
Although we did not perform a quantitative study of the electrical stability, methane admixtures produced fewer discharges in the detector all through the measurements, easing the GEM biasing and the identification of the threshold for discharges. For isobutane admixtures discharges drawing currents above 100 nA were more frequent, even for a stable GEM biasing. For a long-run detector such as the CYGNO-TPC, choosing a gas mixture with low discharge probability is key to preserve detector integrity and the continuity of the experimental campaign.

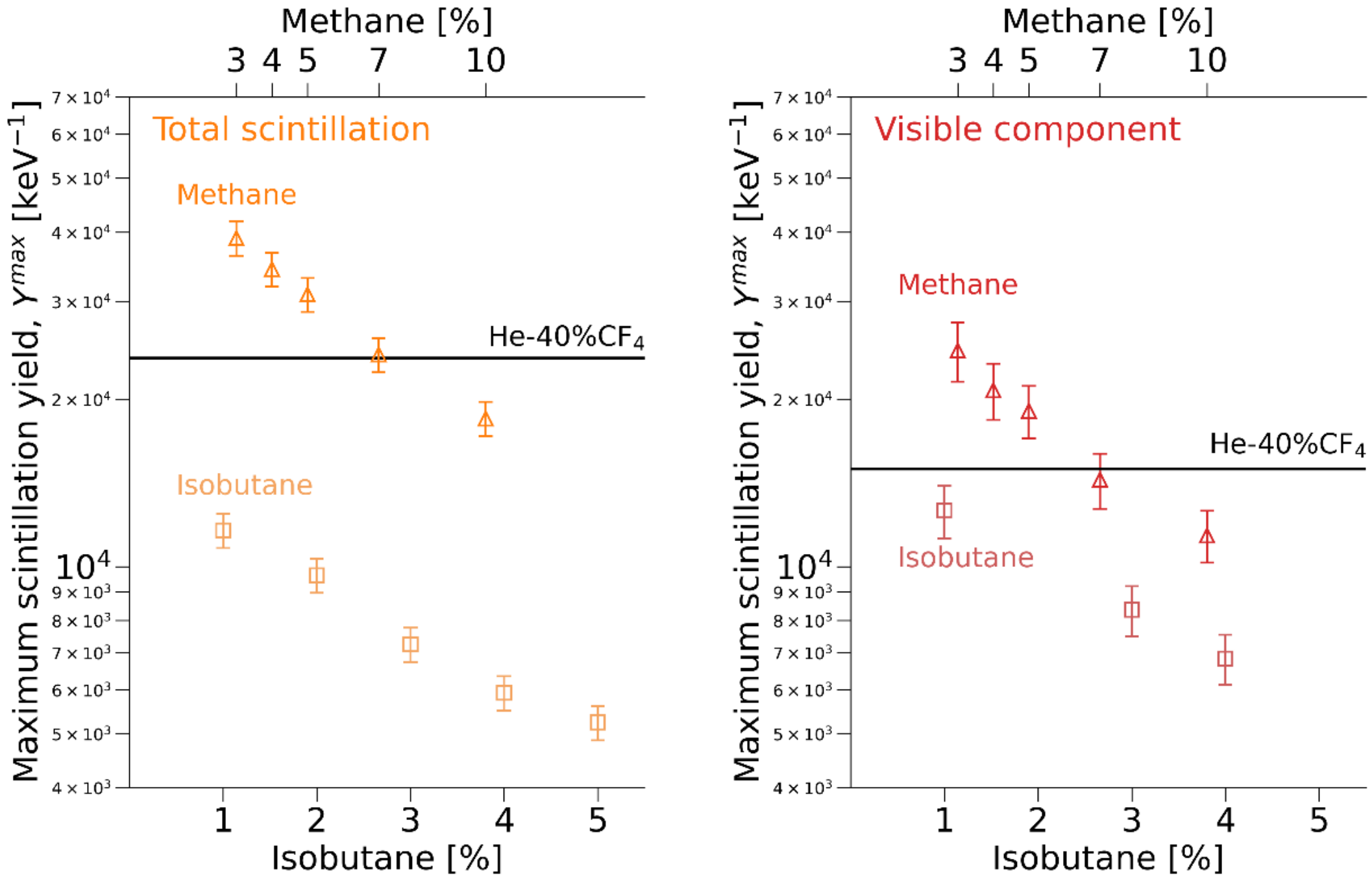

Figure 8: Maximum total scintillation yield (left) and the respective visible component (right) as a function of methane and isobutane addition to He-40%$CF_4$: the hydrocarbon is identified by the corresponding marker shape. Methane and isobutane curves are vertically aligned according to their hydrogen-equivalent concentrations. The horizontal lines correspond to the scintillation yield measured for the base-mixture, He-40%$CF_4$.

* Corresponding author cristinam@uc.pt

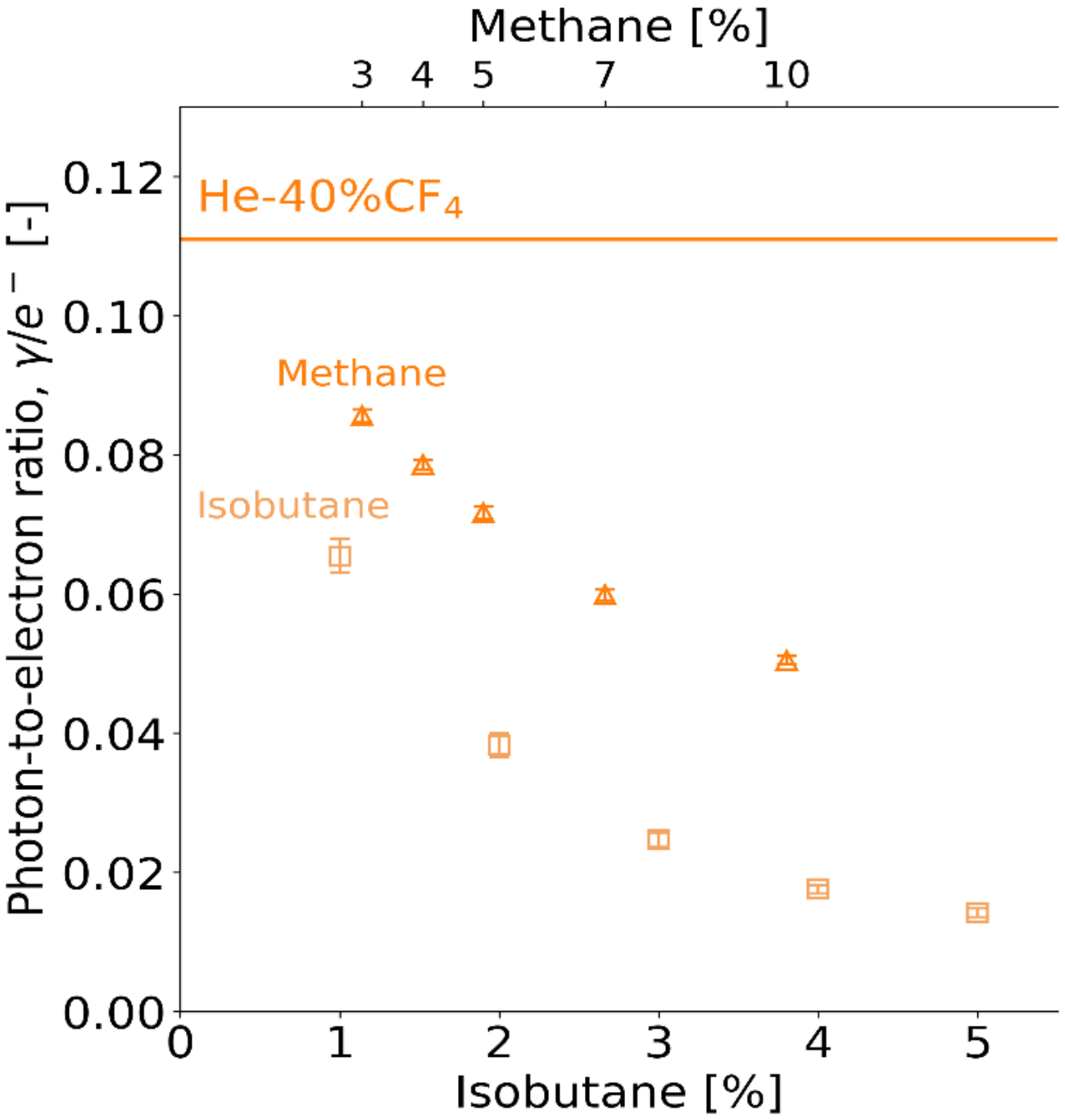

Figure 9: average number of photons per avalanche electron for the total scintillation as a function of the methane and isobutane admixture content to He-40%$CF_4$: the hydrocarbon is identified by the corresponding marker shape. Methane and isobutane curves are vertically aligned according to their hydrogen-equivalent concentrations. The horizontal line corresponds to the yield measured for the base mixture, He-40%$CF_4$.

# Conclusions

The addition of methane to He-$CF_4$ allows to increase the scintillation yield output from GEM avalanches up to 7% of methane content. This happens because the addition of methane increases the electrical stability of the detector and enables higher GEM biasing voltages, which increases the maximum scintillation output relative to He-40%$CF_4$. In opposition, isobutane addition to He-40%$CF_4$ does not increase the threshold for electrical discharges and the scintillation output decreases with increasing isobutane content. For the same hydrogen content, methane admixtures to He-40%$CF_4$ guarantee higher scintillation yields than isobutane, either in terms of maximum achievable scintillation yield, Y, or in terms of the average number of photons emitted per avalanche electron, $Y/e^-$. As in the case of isobutane, the energy resolution does not degrade with the addition of methane.
The obtained results show that methane is a better additive to He-40%$CF_4$ than isobutane in the context of CYGNO: it improves the electrical stability of the detector and increases the maximum attainable scintillation yield, without degrading the corresponding energy resolution. Therefore, CYGNO can improve the detector sensitivity to low WIMP masses while improving its electrical stability and increasing or maintaining the gas scintillation yield by adding up to 7% of methane to the He-40%$CF_4$ base-mixture.

* Corresponding author cristinam@uc.pt

While a detailed quantitative analysis of GEM electrical stability was not conducted, methane admixtures consistently exhibited fewer discharges during measurements. This suggests that methane enhances operational stability, warranting further investigation in future studies.

**Acknowledgements**

R.J.C. Roque acknowledges FCT PhD grant SFRH/BD/143355/2019. This work is supported by 2024.00269.CERN, CERN/FIS-TEC/0038/2021, UIDB/FIS/04559/2025 and UIDP/FIS/04559/2025 (LIBPhys), funded by national funds through FCT/MCTES and co-financed by the European Regional Development Fund (ERDF) through the Portuguese Operational Program for Competitiveness and Internationalization, COMPETE 2020. Part of this project has received funding from the European Union’s Horizon 2020 re-search and innovation programme from the European Research Council (ERC) grant agreement No 818744. This work is partially supported by ICSC — Centro Nazionale di Ricerca in High Performance Computing, Big Data and Quantum Computing, funded by European Union — NextGenerationEU

**References**

[1] E. Aprile et al., Observation and Applications of Single-Electron Charge Signals in the XENON100 Experiment, J. Phys. G 41, 035201 (2014).

[2] D. S. Akerib et al., Extending Light WIMP Searches to Single Scintillation Photons in LUX, Phys. Rev. D 101, 042001 (2020).

[3] C. A. O. Henriques et al., Neutral Bremsstrahlung Emission in Xenon Unveiled, Phys. Rev. X 12 021005 (2022).

[4] M. Adrover et al., Cosmogenic background simulations for the DARWIN observatory at different underground locations, arXiv:2306.16340, 2023.

[5] E. Aprile et al., Material screening and selection for XENON100, Astroparticle Physics 35 (2011) 43-49.

[6] F. Mayet et al., A review of the discovery reach of directional Dark Matter detection,

Phys. Rep. 627 (2016) 1–49.

[7] S. E. Vahsen et al., CYGNUS: Feasibility of a nuclear recoil observatory with directional sensitivity to dark matter and neutrinos, arXiv preprint arXiv:2008.12587 (2020).

[8] Y. Hochberg et al., Directional detection of light dark matter in superconductors, Phys. Rev. D 107 (2023) 076015.

[9] A. Anokhina et al., Directional Observation of Cold Dark Matter Particles (WIMP) in Light Target Experiments, Universe 7 (2021) 215.

[10] J. B. R. Battat et al., Improved sensitivity of the DRIFT-IId directional dark matter experiment using machine learning, JCAP 2021 (2021) 014

[11] F. D. Amaro et al., The CYGNO experiment, Instruments 6 (2022) 6.

[12] F. D. Amaro et al., A 50l CYGNO prototype overground characterization, Eur. J. Phys. C 83 (2023) 946.

* Corresponding author cristinam@uc.pt

[13] E. Baracchini et al., Technical Design Report – TDR CYGNO-04/INITIUM; Technical report, Istituto Nazionale di Fisica Nucleare, INFN-PM-QA-504 Rev. 1.0.1.

[14] V. C. Antochi et al., A GEM-based Optically Readout Time Projection Chamber for charged particle tracking, arXiv preprint arXiv:2005.12272 (2005).

[15] V. C. Antochi et al., Combined readout of a triple-GEM detector, J. Instrum. 13 (2018) P05001.

[16] M. M. F. R. Fraga et al., The gem scintillation in He-CF4, Ar-CF4, Ar-TEA and Xe-TEA mixtures, Nucl. Instrum. Meth. A 504 (2003) 88–92.

[17] F. D. Amaro et al., Secondary scintillation yield from GEM electron avalanches in He-CF4 and He-CF4-isobutane for CYGNO — Directional Dark Matter search with an optical TPC, Phys. Lett. B 855 (2024) 138759.

[18] https://www.advancedphotonix.com/, 1240 Avenida Acaso, Camarillo, CA 93012 USA.

[19] https://www.crystran.com/optical-materials/optical-glass, CRYSTRAN, Optical Glass (N-BK7 types), CRYSTRAN (2012).

[20] C. M. B. Monteiro et al., Secondary scintillation yield in pure xenon, J. Instrum. 2 (2007) P05001.

[21] C. M. B. Monteiro et al., Secondary scintillation yield in pure argon, Phys. Lett. B 668 (2008) 167–170.

[22] C. M. B. Monteiro et al., Secondary scintillation yield from gaseous micropattern electron multipliers in direct dark matter detection, Phys. Lett. B 677 (2009) 133–138.

[23] C. M. B. Monteiro et al., Secondary scintillation yield from GEM and THGEM gaseous electron multipliers for direct dark matter search, Phys. Lett. B 714 (2012) 18–23.

[24] G. F. Knoll, Radiation detection and measurement John Wiley & Sons, Vol. 65, John Wiley & Sons, 2000.

[25] NIST, Ionization Energy Evaluation, NIST Chemistry WebBook, SRD 69.

[26] A. Janeco et al., Electron Kinetics in He/CH4/CO2 Mixtures Used for Methane Conversion, The Journal of Physical Chemistry C, 119(1):109–120, 2015.

[27] C. A. O. Henriques et al., Electroluminescence TPCs at the thermal diffusion limit, JHEP 01 (2019) 027.

* Corresponding author cristinam@uc.pt